\documentclass[aps,prl,twoside,twocolumn,nofootinbib,10pt,showpacs,floatfix]{revtex4-1}
\usepackage{amsmath,amssymb}
\usepackage{graphicx,bm}
\usepackage{slashed}
\usepackage{epstopdf}
\usepackage{ulem} 
\usepackage[usenames]{color}
\usepackage{float}
\usepackage{braket}
\usepackage{hyperref}
\usepackage{subfigure}
\usepackage{subfigure}
\usepackage{rotating}
\usepackage{color}
\usepackage{multirow}
\usepackage{dcolumn}
\usepackage{overpic}
\usepackage{booktabs}
\usepackage{makecell}
\usepackage{diagbox}
\renewcommand\sout{\bgroup \color{red} \ULdepth=-.5ex \ULset}

\newsavebox{\tablebox}

\begin{document}
	\title{Complementary $CP$ violation induced by $T$-odd and $T$-even correlations}
	\author{Jian-Peng Wang$^1$}\email{wangjp20@lzu.edu.cn}
	\author{Qin Qin$^2$}\email{qqin@hust.edu.cn, corresponding author}
	\author{Fu-Sheng Yu$^{1}$}\email{yufsh@lzu.edu.cn, corresponding author}
	\affiliation{
		$^1$MOE Frontiers Science Center for Rare Isotopes, and School of Nuclear Science and Technology, Lanzhou University, Lanzhou 730000, China\\
		$^2$School of physics, Huazhong University of Science and Technology, Wuhan 430074, China\\
	}
	\begin{abstract}
	In this letter, we propose a novel approach to concurrently measure the complementary $CP$ violation observables induced by $T$-odd correlations and their corresponding $T$-even counterparts, where $T$ represents time reversal. Our analysis demonstrates that $T$-odd and -even correlations, when satisfying specific conditions, result in cosine and sine strong phase dependencies of the corresponding $CP$ violation, respectively. Additionally, we identify pairs of these $CP$ violation observables in hadron decays depend on precisely the same strong phases within the helicity amplitude scheme. This complementarity effectively reduces the strong phase reliance in the study of $CP$ violation, while also mitigating the risk of suppressed $CP$ violation due to exceptionally small strong phases. Furthermore, our proposal holds potential for uncovering $CP$ violation in baryon decays that have not yet been observed in experiments.
	\end{abstract}
	\maketitle
	
	\noindent{\it Introduction.--}
	Understanding the asymmetry between baryons and anti-baryons in the universe is a significant challenge in modern particle physics and cosmology. This puzzle can be addressed by satisfying three conditions known as the Sakharov criteria~\cite{Sakharov:1967dj}: baryon number violation, $C$ and $CP$ violation (CPV), and departure from equilibrium. In the Standard Model (SM) of particle physics, the only confirmed CPV source is the weak phase in the quark mixing matrix, as proposed by the Kobayashi-Maskawa (KM) mechanism~\cite{Kobayashi:1973fv}. However, the level of CPV in the SM is not adequate to account for the matter-dominated universe as observed \cite{Planck:2015fie}, suggesting the presence of additional CPV sources. Furthermore, precise CPV measurements are crucial for determining the elements of the KM matrix, which is essential for testing the unitarity of the KM matrix required by the SM. Therefore, CPV serves as a promising avenue to exploring new physics beyond the SM.
	
In flavor physics, extensive research has been conducted on CPV in meson decays and mixing~\cite{Christenson:1964fg,BaBar:2001pki,Belle:2001zzw,LHCb:2019hro,ParticleDataGroup:2022pth}. Notable achievements, such as the discovery of CPV in $B^0\to J/\psi K_S$~\cite{BaBar:2001pki,Belle:2001zzw}, have confirmed the validity of the KM mechanism. However, despite the accumulation of more data and higher-order calculations, precision tests of CPV observables in most decay channels still face challenges in reconciling theory and experiment, hindering the search for non-standard dynamics. This difficulty is particularly evident in the case of direct $CP$ asymmetry, which is proportional to the sine of the strong phase. Theoretical calculations of strong phases often introduce significant uncertainties. In order to tackle this issue, new CPV observables have emerged, such as mixing involved CPV~\cite{Yu:2017oky,Shen:2023nuw}, partial-wave CPV~\cite{Zhang:2021zhr,Zhang:2021fdd}, triple-product CPV~\cite{Chiang:1999qn,Valencia:1988it}, and others. Some of these observables exhibit a cosine dependence on strong phases, including CPV induced by triple products, Lee-Yang asymmetries, and more general $T$-odd correlations~\cite{Valencia:1988it,Donoghue:1985ww,Donoghue:1986hh,Chiang:1999qn,Lee:1956qn,Durieux:2015zwa,Gronau:2011cf,Datta:2003mj,Bevan:2014nva,Ajaltouni:2004zu,Leitner:2006sc,Gronau:2015gha,Geng:2021sxe}. This characteristic potentially allows for the cancellation of strong phase dependence if two CPV observables depend on the sine and cosine of exactly the same strong phase~\cite{Valencia:1988it,Donoghue:1985ww,Donoghue:1986hh,Durieux:2015zwa}. We refer to this phenomenon as true complementarity.\footnote{Conversely, if two CPV observables are proportional to sine and cosine of different strong phases, one cannot conclude that they are complementary to each other.} For instance, a CPV observable constructed from $\alpha$ and $\beta$ can be approximately independent of strong phases~\cite{BESIII:2021ypr}.  Despite these advancements, there remain unresolved mysteries in the conventional discussions of $T$-odd correlations and  complementarity. These mysteries can be summarized as follows: 
	\begin{itemize}
		\item What is the underlying reason for the cosine dependence of strong phases in $T$-odd correlation induced CPV?
		\item Are true complementarities widespread, and if so, how can we identify them? 
	\end{itemize}
	
By this work, we clarify the aforementioned questions by offering two rigorous proofs: {\bf (1)} We provide a strict proof showing that the $T$-odd correlation induced CPV indicates a cosine dependence on the strong phase under certain conditions, and that the corresponding $T$-even correlation induced CPV indicates a sine dependence; {\bf (2)} We present the criteria for true complementarity between pairs of $T$-odd and -even CPV observables in two-body decays with helicity amplitude framework. Based on the proofs, we propose the feasibility of simultaneously measuring a pair of CPV observables that exhibit dependencies on $\sin\delta_{s}$ and $\cos\delta_{s}$ relative to the same strong phase difference $\delta_{s}$. Our proof will also provide a systematic way to find this type of complementary observation, and thus can lead to a blanket search for the complementary $T$-odd and -even CPV observables.
	
	We also emphasize that the complementarity holds promise for the search for baryonic CPV, which has yet to be discovered. Complementary CPV observables provide an avenue for detecting CPV that is suppressed by small strong phases~\cite{BESIII:2018cnd,BESIII:2021ypr}, and regardless of the strong phase value, since either $\sin\delta_s$ or $\cos\delta_s$ exceeds $\sqrt{2}/2$. Finally, we demonstrate the feasibility of our proposal experimentally through a specific example of $\Lambda_{b}^{0}\to N^{*}(1520)K^{*}(892)$ with $N^{*}(1520)\to p\pi$ and $K^{*}(892)\to K\pi$, and analyze its potential applications in other decays involving baryons.
	
\noindent{\it Strong phase dependence--}As the first step, we prove that the CPV $a^{Q_{-}}_{\rm CP}$ induced by a subset of $T$-odd correlations $Q_-$ are proportional to the cosine of the involved strong phase differences, $\cos\delta_s$. The $T$-odd property of $Q_-$ indicates its transformation under the time reversal $\mathcal{T}$ as 
	\begin{eqnarray}
		\mathcal{T}Q_- = -Q_-\mathcal{T} \;.
	\end{eqnarray} 
	It is important to note that not all $Q_-$ can generate $CP$ asymmetries proportional to $\cos\delta_s$ (see {\it e.g.}~\cite{Bevan:2014nva}). We propose that a qualified $Q_-$ satisfies the following conditions: {\bf (i)} In the Hilbert space of the final states of a physical process of interest, with a properly chosen basis \{$|\psi_n\rangle$, n =1,2,...\}, there exists a unitary transformation $\mathcal{U}$ that transforms $\mathcal{T}|\psi_n\rangle$ back to $|\psi_n \rangle$ up to a universal phase factor, {\it i.e.}, $\mathcal{U}\mathcal{T}|\psi_n\rangle  = e^{i\alpha}|\psi_n\rangle$; {\bf(ii)} $Q_-$ is symmetric under this unitary transformation, {\it i.e.} $\mathcal{U}Q_-\mathcal{U}^\dagger = Q_-$.  The proof of $a^{Q_-}_{\rm CP}$ being proportional to $\cos\delta_s$ is as follows. 
	
	The $Q_-$ expectation value of the final state $|f\rangle\equiv S|i\rangle$ of a process, with $S$ being the S-matrix operator, can be expressed in terms of the transition amplitudes from the initial state to basis vectors $A_n\equiv  \langle \psi_n |  S |i \rangle$, as
	\begin{eqnarray}\label{eq:qexp}
		\langle f| Q_- | f \rangle &=& \langle i | S^\dagger Q_- S | i \rangle  \nonumber \\
		&=& \sum_{m,n} \langle \psi_i | S^\dagger |\psi_m\rangle \langle \psi_m|  Q_-  | \psi_n \rangle \langle \psi_n |  S |\psi_i \rangle  \nonumber \\
		&=& \sum_{m,n} A_m^* A_n \langle \psi_m|  Q_-  | \psi_n \rangle \; . 
	\end{eqnarray}
	The dynamics are now coded in $A_n$'s, and $\langle \psi_m|  Q_-  | \psi_n \rangle$'s only consist of kinematics. 
	Then it can be shown that the matrix element $\langle \psi_m|  Q_-  | \psi_n \rangle$ is purely imaginary by 
	\begin{eqnarray}\label{eq:qmn}
		\langle \psi_m|  Q_-  | \psi_n \rangle 
		&=&\langle \psi_m| \mathcal{T}^\dagger  \mathcal{T} \; Q_-  | \psi_n \rangle^* \nonumber \\
		&=& - \langle \psi_m| \mathcal{T}^\dagger  Q_- \mathcal{T} | \psi_n \rangle^* \nonumber \\
		&=& - \langle \psi_{m}| \mathcal{T}^\dagger \; \mathcal{U}^\dagger \mathcal{U}\; Q_- \; \mathcal{U}^\dagger \mathcal{U} \; \mathcal{T} | \psi_n \rangle^* \nonumber \\
		&=& - \langle \psi_{m}| \mathcal{T}^\dagger \mathcal{U}^\dagger \; Q_- \; \mathcal{U} \mathcal{T} | \psi_n \rangle^* \nonumber \\
		&=& - \langle \psi_{m}|  Q_-  | \psi_n \rangle^* \; ,
	\end{eqnarray}
	where in the first step the anti-unitarity of $\mathcal{T}$ is used. Consequently, only the imaginary part of the amplitude interference $\mathrm{Im}(A_m^* A_n)$ contributes, because $\langle f| Q_- | f \rangle$ must be real. This conclusion holds true for both perturbative and non-perturbative dynamics, and for diverse physical systems such as beauty, charm, strange, top and even Higgs physics. 
	
	The $CP$ asymmetry  induced by a $T$-odd correlation $Q_{-}$ is defined as
	\begin{eqnarray}\label{eq:CP}
		a^{Q_{-}}_{\rm CP}\equiv {\langle f | Q_-  | f \rangle  - \langle \bar{f} |   \bar{Q}_-  | \bar{f} \rangle } \; ,
	\end{eqnarray}
	where $\ket{\bar{f}}\equiv S(CP)\ket{i}$  and $\bar{Q}_-\equiv (CP){Q}_-(CP)^{-1}$. 
	By inserting a complete basis of  $|\psi_n\rangle$ and $|\bar{\psi}_n\rangle \equiv CP|\psi_n\rangle$, we obtain
	\begin{eqnarray}\label{eq5}
		a^{Q_{-}}_{\rm CP} &\propto & \sum_{m,n} i \; \mathrm{Im}(A_m^* A_n - \bar{A}_m^* \bar{A}_n )  \langle \psi_{m}|  Q_-  | \psi_n \rangle \; ,
	\end{eqnarray}
	where the relation $\langle \psi_{m}|  Q_-  | \psi_n \rangle = \langle \bar{\psi}_{m}|  \bar{Q}_-  | \bar{\psi}_n \rangle$ independent of dynamics has been utilized. In quark-flavor processes whose CPV is induced by the KM mechanism, the imaginary $CP$ differences $\mathrm{Im}(A_m^* A_n - \bar{A}_m^* \bar{A}_n)$ must be proportional to the sine of the weak phase difference $\sin\delta_w$, and hence the cosine of the relevant strong phase difference $\cos\delta_s$. 
	
	Analogously, if a $T$-even correlation $Q_+$ satisfies conditions {\bf(i)} and {\bf(ii)}, the right-hand side of \eqref{eq:qmn} flips the sign, such that the $Q_+$ expectation depends on the real part of amplitude interferences and, of course, on the possible modulo terms. Therefore, its induced CPV will be proportional to the sine of the strong phase difference. 
	In fact, direct CPV is induced by a $T$-even correlation, which can be defined by $|f_d\rangle \langle f_d|$ with $|f_d\rangle$ the desired final state, so they have the sine dependence on $\delta_s$. 
	If a pair of $\langle{Q_-}\rangle$ and $\langle{Q_+}\rangle$ pick the $\mathrm{Im}(A_m^* A_n)$ and $\mathrm{Re}(A_m^* A_n)$ contributions, respectively, with the same weights, we will prove that they give rise to $CP$ asymmetries proportional to the cosine and sine of the same strong phase in the subsequent section. From this perspective, they exhibit an exact complementary relationship with each other.

It is important to note that the above proposition is not limited to time reversal but applies universally to any anti-unitary transformation, such as the combined transformation of spatial and time reversals $P\mathcal{T}$. In addition, the condition {\bf(i)} can be slightly relaxed: it is sufficient that $\langle \psi_{m}| \mathcal{T}^\dagger \mathcal{U}^\dagger \; Q_- \; \mathcal{U} \mathcal{T} | \psi_n \rangle = \langle \psi_{m}| Q_-  | \psi_n \rangle$ instead of requiring $\mathcal{U}\mathcal{T}|\psi_n\rangle  = e^{i\alpha}|\psi_n\rangle$.
	
Our prescription can be easily applied to two-body hadron decays involving at least two non-zero spin particles. The $T$-odd correlation $Q_-$ can be selected as an odd-multiple-product of spin and momentum vectors of the particles involved, such as the triple-product $(\vec{s}_{1}\times\vec{s}_2)\cdot\vec{p}$, where the particle spins are defined in the rest frame of each respective particle. Correspondingly, the unitary transformation $\mathcal{U}$ is chosen as the spatial rotation, and the basis vectors $\ket{\psi_{n}}$ are selected as the helicity eigenstates. The final-state helicity eigenstates are denoted by $|J,M;\lambda_1,\lambda_2\rangle$, where $J$ is the final-state angular momentum, $M$ is its $z$-direction component, which are determined by the initial state, and $\lambda_1$ and $\lambda_2$ are the helicities of the two final-state particles. Following the convention of~\cite{Jacob:1959at}, the time reversal $\mathcal{T}$ and the rotation about the $y$-axis by $\pi$, $\mathcal{U}=e^{-i\pi J_y}$, both transform $|J,M;\lambda_1,\lambda_2\rangle$ to $(-1)^{J-M}|J,-M;\lambda_1,\lambda_2\rangle$. Therefore, the condition {\bf(i)} is satisfied, with 
	\begin{eqnarray}
		\mathcal{U}\mathcal{T} |J,M;\lambda_1,\lambda_2\rangle = (-1)^{2J}|J,M;\lambda_1,\lambda_2\rangle \; .
	\end{eqnarray}
Furthermore, the triple-products, being spatial-SO(3) scalars, remain invariant under spatial rotations, thus fulfilling condition {\bf(ii)}. Subsequently, we will delve into further details regarding this type of decay processes, elucidating the genuine complementarity of the CP violation observables involved. \footnote{It is worth noting that the $T$-odd triple product $(\vec{p}_{1}\times\vec{p}_2)\cdot\vec{p}_3$ consists of three momentum in four-body decays can not satisfy the conditions {\bf(i)} and {\bf(ii)} simultaneously. The $\mathcal{T}$ transformation flips all the particle momenta, so the condition {\bf(i)} requires that $\mathcal{U}$ flips the momenta back. Then, we must have $\mathcal{U}\; (\vec{p}_{1}\times\vec{p}_2)\cdot\vec{p}_3 \;\mathcal{U}^\dagger = - (\vec{p}_{1}\times\vec{p}_2)\cdot\vec{p}_3$ and thus the condition {\bf(ii)} is not satisfied. Therefore, the corresponding CPV is not necessarily proportional to $\cos\delta_s$~\cite{Bevan:2014nva}.}
	
	\noindent{\it Criteria for complementary observable--}As demonstrated earlier, $T$-odd and -even correlations satisfying conditions {\bf(i)} and {\bf(ii)} induce CPV observables with cosine and sine dependences on strong phases, respectively. However, a critical question remains as to how to determine whether two observations are exactly complementary. Providing a general answer to this question would be quite challenging. Instead, we will limit ourselves to two-body decays and select the final-state bases to be the helicity eigenstates~\cite{Geng:2021sxe}. In this context, we introduce a criterion within the helicity framework.
	
	\begin{itemize}
		\item {\bf Criterion:} If two observables exhibit dependencies on the real and imaginary parts of the same interference term under the helicity amplitude scheme, then they will induce exactly complementary CPV observables.
	\end{itemize}
	{\bf Proof}: In the helicity bases, the expression of $\langle Q_-\rangle$ \eqref{eq:qexp} is composed by helicity amplitude interferences and $\langle Q_+\rangle$ is analogous. Consider the simplest case where two operators $\mathcal{O}_{+}$ and $\mathcal{O}_{-}$ have expectations given by 
	\begin{equation}
		\begin{aligned}    \langle\mathcal{O}_{+}\rangle&=\mathcal{R}e(\mathcal{H}_{\lambda_{i},\lambda_{j}}\mathcal{H}^{*}_{\lambda_{m},\lambda_{n}}+\mathcal{H}_{-\lambda_{i},-\lambda_{j}}\mathcal{H}^{*}_{-\lambda_{m},-\lambda_{n}}) \; ,\\  
			\langle\mathcal{O}_{-}\rangle&=\mathcal{I}m(\mathcal{H}_{\lambda_{i},\lambda_{j}}\mathcal{H}^{*}_{\lambda_{m},\lambda_{n}}+\mathcal{H}_{-\lambda_{i},-\lambda_{j}}\mathcal{H}^{*}_{-\lambda_{m},-\lambda_{n}}) \; , \\
		\end{aligned}
	\end{equation}
	where $\lambda_{i,j},\lambda_{m,n}$ are general helicity indices of the final-state particles. This can be fulfilled when the operators have only nonzero matrix elements $\langle\lambda_m,\lambda_n | \mathcal{O}_\pm | \lambda_i,\lambda_j \rangle $ and $\langle -\lambda_m,-\lambda_n | \mathcal{O}_\pm | -\lambda_i,-\lambda_j \rangle $. 
	Note that both $\langle\mathcal{O}_{+}\rangle$ and $\langle\mathcal{O}_{-}\rangle$ comprise two terms linked by the parity transformation. This choice is reasonable because observables that we are interested in invariably manifest specific symmetries under spatial inversion, such as triple products~\cite{Valencia:1988it,Bevan:2014nva} and asymmetry parameters~\cite{Lee:1956qn}. Here, one can check both of them are parity even. This proof also remains applicable for the opposite case. The $CP$ asymmetries induced by $\mathcal{O}_{\pm}$ are defined by
	\begin{equation}
		\begin{aligned}
			a^{\mathcal{O}_{+}}_{CP}={\langle\mathcal{O}_{+}\rangle- \langle\bar{\mathcal{O}}_{+}\rangle},~~ a^{\mathcal{O}_{-}}_{CP}={\langle\mathcal{O}_{-}\rangle-\langle\bar{\mathcal{O}}_{-}\rangle}
		\end{aligned}
	\end{equation}
	where $\langle\bar{\mathcal{O}}_{\pm}\rangle$ are the corresponding charge conjugations. They can be further normalized to make them dimensionless.

	A helicity amplitude can be decomposed into tree and penguin contributions as 
	\begin{equation}
		\mathcal{H}_{\lambda_{i},\lambda_{j}}=H^{t}_{{i},{j}} e^{i\phi_{t}}e^{i\delta^{t}_{i,j}}+H^{p}_{{i},{j}} e^{i\phi_{p}}e^{i\delta^{p}_{i,j}} \; ,
	\end{equation}
	where $H^{t (p)}_{\lambda_{i},\lambda_{j}},\delta^{t (p)}_{i,j},\phi_{t (p)}$ are the magnitude, strong and weak phases of the tree (penguin) amplitude, respectively. Its $CP$ conjugation partner $\bar{\mathcal{H}}_{\lambda_{i},\lambda_{j}}$ can be correspondingly expressed as 
	\begin{equation}
		\bar{\mathcal{H}}_{-\lambda_{i},-\lambda_{j}}=H^{t}_{{i},{j}} e^{-i\phi_{t}}e^{i\delta^{t}_{i,j}}+H^{p}_{{i},{j}} e^{-i\phi_{p}}e^{i\delta^{p}_{i,j}} \; ,
	\end{equation}
	by flipping the weak phase signs. It can be different by an overall minus sign depending on the $CP$ transformation conventions of the initial and final states, which does not change the physics. The similar relation holds between their parity partners. This leads to a comprehensive complementary observation.
	
	\begin{equation}\label{eq9}
		\begin{aligned}
			a^{\mathcal{O}_{+}}_{CP} \propto &[-{H}^{t}_{i,j} {H}^p_{m,n} \sin(\delta^{t}_{i,j}-\delta^{p}_{m,n})\\
			&+ {H}^{p}_{i,j} {H}^t_{m,n} \sin(\delta^{p}_{i,j}-\delta^{t}_{m,n})]\sin\Delta\phi\\
			&+(i,j,m,n\to -i,-j,-m,-n) \; , \\
			a^{\mathcal{O}_{-}}_{CP} \propto &[- {H}^{t}_{i,j} {H}^p_{m,n} \cos(\delta^{t}_{i,j}-\delta^{p}_{m,n})\\
			&+ {H}^{p}_{i,j} {H}^t_{m,n} \cos(\delta^{p}_{i,j}-\delta^{t}_{m,n})]\sin\Delta\phi\\
			&+(i,j,m,n\to -i,-j,-m,-n) \; , \\
		\end{aligned}
	\end{equation}
	where $\Delta\phi \equiv \phi_t-\phi_p$. It can be observed that $a^{\mathcal{O}_{+}}_{CP},a^{\mathcal{O}_{-}}_{CP}$ are dependent on the identical set of strong phase differences, and thus exactly complementary to each other. This establishes the complementarity under the helicity scheme. It is crucial to highlight that the complementarity exists between $a^{\mathcal{O}_{+}}_{CP}$ and $a^{\mathcal{O}_{-}}_{CP}$, rather than between $a^{\mathcal{O}_{-}}_{CP}$ and the direct $CP$ asymmetry. The direct $CP$ asymmetry characterizes the difference between the total widths $\Gamma$ and $\bar{\Gamma}$ consisting of the modulo squared of distinct helicity configurations, while the $T$-odd $CP$ asymmetry, as in~\eqref{eq5}, consists of interference terms, so they rely on different strong phases. 
	
The discussions presented above are focused on two-body decays. However, the situation becomes more complex in the case of multibody systems due to the presence of intricate intermediate resonances. Consequently, the applicability of the aforementioned proof might be compromised in such scenarios. Nevertheless, it is worth noting that the amplitude $\mathcal{H}_{\lambda_{i},\lambda_{j}}$ can be directly extracted in experiments by employing the partial wave analysis method in multibody decays~\cite{BESIII:2022udq,BESIII:2022bvv}. In this context, our proposal retains its value and practicality, providing a useful framework for analyzing and interpreting experimental results in multibody systems.
	
\begin{figure}[!]
		\centering
		\includegraphics[width=0.48\textwidth]{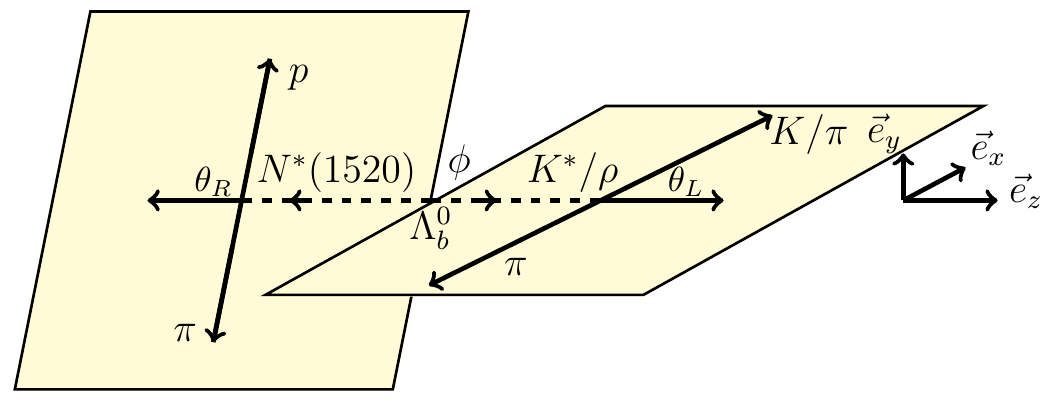}
		\caption{The figure illustrates the decay process of $\Lambda^{0}_{b}\to N^{*}(1520)K^{*}/\rho$ with $N^{*}(1520)\to p\pi^{-},K^{*}(\rho)\to K(\pi)\pi$ decay. The angles $\phi$, $\theta_{L}$ and $\theta_{R}$ are defined in the rest frames of $\Lambda_b$, $K^{*}(\rho)$ and $N^{*}(1520)$, respectively.}\label{fig}
	\end{figure}

	\noindent{\it Examples in baryon sector--} 
Our proposal has a wide range of applications in decay processes involving baryons. Given that the helicity information of the final-state particles undergoing subsequent decays is manifested in the angular distribution of their decay products, the helicity amplitudes of a cascade decay can be derived from its angular distribution. As an illustration, we analyze the decay channel $\Lambda^{0}_{b}\to N^{*}(1520)K^{*}$ with $N^{*}(1520)\to p\pi, K^{*}(892)\to K\pi$~\cite{LHCb:2019jyj}. The results apply directly to the similar $\Lambda^{0}_{b}\to N^{*}(1520)\rho$ decay with $\rho\to \pi^+\pi^-$. With unpolarized $\Lambda^{0}_{b}$, the complementary part of angular distribution is formulated as
	\begin{equation}\label{eq:angularNstarKstar}
		\begin{aligned}
			&\frac{d\Gamma}{d\mathrm{c}_{1} \, d\mathrm{c}_{2}\, d\phi_{R}}\ni \\
			&-\frac{\mathrm{s}_{L}^2 \mathrm{s}_{R}^2}{\sqrt{3}} \mathrm{Im}[\mathcal{H}_{+1,+\frac{3}{2}}\mathcal{H}_{-1,-\frac{1}{2}}^{*} +\mathcal{H}_{+1,+\frac{1}{2}}\mathcal{H}_{-1,-\frac{3}{2}}^{*}]\sin2\phi\\
			&+\frac{\mathrm{s}_{L}^2 \mathrm{s}_{R}^2}{\sqrt{3}} \mathrm{Re}[\mathcal{H}_{+1,+\frac{3}{2}}\mathcal{H}_{-1,-\frac{1}{2}}^{*} + \mathcal{H}_{+1,+\frac{1}{2}}\mathcal{H}_{-1,-\frac{3}{2}}^{*}] \cos2\phi\\&
			-\frac{4\mathrm{s}_{L}\mathrm{c}_{L}\mathrm{s}_{R}\mathrm{c}_{R}}{\sqrt{6}} \mathrm{Im}[\mathcal{H}_{+1,+\frac{3}{2}}\mathcal{H}_{0,+\frac{1}{2}}^{*} + \mathcal{H}_{0,-\frac{1}{2}}\mathcal{H}_{-1,-\frac{3}{2}}^{*}] \sin\phi\\&
			+\frac{4\mathrm{s}_{L}\mathrm{c}_{L}\mathrm{s}_{R}\mathrm{c}_{R}}{\sqrt{6}} \mathrm{Re}[\mathcal{H}_{+1,+\frac{3}{2}}\mathcal{H}_{0,+\frac{1}{2}}^{*} + \mathcal{H}_{0,-\frac{1}{2}}\mathcal{H}_{-1,-\frac{3}{2}}^{*}] \cos\phi \;  ,
		\end{aligned}
	\end{equation}
where $\mathrm{s}_{L,R} = \sin\theta_{L,R}$ and $\mathrm{c}_{L,R} = \cos\theta_{L,R}$. The angular variables $\theta_{L,R}$ represent the polar angles of the proton and $K$ meson in the rest frame of $N^{*}(1520)$ and $K^{*}$, respectively, and $\phi$ denotes the angle between the decay planes of $N^{*}(1520)$ and $K^{*}$, as depicted in FIG~\ref{fig}. The amplitudes $\mathcal{H}_{\lambda_{1},\lambda_{2}}$ parameterize the dynamics of the $\Lambda_{b}^{0}\to N^{*}(1520)K^{*}$ decay, with $\lambda_{1},\lambda_{2}$ being helicity symbols of $K^{*}$ and $N^{*}(1520)$, respectively.
	 Here, we define two $T$-odd parameters with respect to $\sin\phi_{R}$ and $\sin2\phi_{R}$ as $\mathcal{A}_{T,1} \equiv \mathrm{Im} (\mathcal{H}_{+1,+{3\over2}} \mathcal{H}^*_{0,+{1\over2}} + \mathcal{H}_{-1,-{3\over2}}^* \mathcal{H}_{0,-{1\over2}})$ and $\mathcal{A}_{T,2} \equiv \mathrm{Im} \;( \mathcal{H}_{+1,+{3\over2}} \mathcal{H}^*_{-1,-{1\over2}} + \mathcal{H}_{-1,-{3\over2}}^* \mathcal{H}_{+1,+{1\over2}} )$. These parameters can be obtained by integrating the differential decay width using the expressions 
	\begin{eqnarray}
		\mathcal{A}_{T,i} \propto  \int \frac{d\Gamma}{d\mathrm{c}_{L} \, d\mathrm{c}_{R}\, d\phi_{R}} W_i \; d\mathrm{c}_{L} \, d\mathrm{c}_{R}\, d\phi, 
	\end{eqnarray}
	with the weight functions $W_1 = \sin\phi \, \mathrm{c}_{L} \, \mathrm{c}_{R}$ and $W_2 = \sin2\phi\, \mathrm{s}_{L} \, \mathrm{s}_{R}$. The expectations of corresponding $T$-even correlations $\mathcal{B}_{T,i}$ are also defined through the angular distribution with respect to $\cos\phi_{R}$ and $\cos2\phi_{R}$, $\mathcal{B}_{T,1} \equiv \mathrm{Re} \;(\mathcal{H}_{+1,+{3\over2}} \mathcal{H}^*_{0,+{1\over2}} + \mathcal{H}_{-1,-{3\over2}}^* \mathcal{H}_{0,-{1\over2}}),~\mathcal{B}_{T,2} \equiv \mathrm{Re} \;( \mathcal{H}_{+1,+{3\over2}} \mathcal{H}^*_{-1,-{1\over2}} + \mathcal{H}_{-1,-{3\over2}}^* \mathcal{H}_{+1,+{1\over2}} )$, and can be analogously extracted. Subsequently, the induced $CP$ asymmetries are given by the differences between $\mathcal{A(B)}_{T,i}$ and their charge conjugations
	\begin{eqnarray}
		a_{CP}^{i} = {\mathcal{A}_{T,i} - \bar{\mathcal{A}}_{T,i} } \, ~~b_{CP}^{i} = {\mathcal{B}_{T,i} - \bar{\mathcal{B}}_{T,i} } \, .
	\end{eqnarray}
	It is important to emphasize that $a_{CP}^{i},b^{i}_{CP}$  are proportional to the cosine and sine of identical strong phases, as previously demonstrated.
	
	If an initially polarized baryon in a decay is considered, the angular analysis becomes more intricate, leading to the emergence of more complementary $CP$ asymmetries ~\cite{Durieux:2016nqr,Geng:2021sxe}. However, the polarization of $b$-baryons produced in $pp$ collision at LHC is negligible, rendering it ineffective for phenomenological analysis~\cite{ATLAS:2014swk,LHCb:2013hzx,CMS:2018wjk,LHCb:2020iux}. Fortunately, the charm and strange baryons produced at lepton colliders are found to have sizable polarization~\cite{BESIII:2018cnd,BESIII:2019odb,Dharmaratna:1996xd,Falk:1993rf,Fanti:1998px}, allowing for a more comprehensive angular analysis.
	
	We anticipate that our proposal will offer significant advantages for the search for CPV in baryonic processes. In addition to the previously analyzed channels, we recommend conducting analogous angular distribution analyses for other other $b$-baryon decays in experiments, such as $\Lambda^{0}_{b}\to \Lambda^{*}(1520)\rho/\phi,\Lambda^{0}_{b}\to pa_{1}$, and so on~\cite{Durieux:2016nqr,Geng:2021sxe,Rui:2022jff,LHCb:2016hwg,LHCb:2018fpt,LHCb:2019oke,LHCb:2019jyj}. Furthermore, similar complementary $CP$ asymmetries are expected in baryonic meson decays, such as $B^{0}\to\Lambda^{+}_{c}\bar{\Lambda}^{-}_{c},\bar{\Xi}^{-}_{c}\Lambda^{+}_{c},\Lambda\bar{\Lambda}$, warranting further investigation.
	
	\noindent{\it Summary.--}In this study, we have addressed the questions surrounding conventional $T$-odd correlation discussions by providing two rigorous proofs. Our findings reveal that the flavor CPV observables induced by $T$-odd correlations, under specific conditions, are directly proportional to the cosine of strong phases, while the corresponding $T$-even correlations give rise to strong-phase-sine CPV. Furthermore, within the helicity representation framework, we have demonstrated a true complementary dependence of strong phases between CPV observables induced by pairs of $T$-odd and -even correlations, whose expectations are proportional to the imaginary and real parts of the same helicity amplitude interferences. This provides a strong basis and could be effectively utilized to reduce the strong phase reliance of CPV, as well as to investigate $CP$ asymmetries in baryon decays. Detailed analysis of practical examples involving $b$-baryon decays demonstrates that the proposed CPV observables can be extracted by measuring the angular distribution of the decay products. As the amplitude analysis method continues to develop and be applied in experimental settings, these complementary forms of CPV will increasingly shape the landscape of future research endeavors.

	\noindent{\it Acknowledgement.--}This work is supported in part by the National Natural Science Foundation of China under Grants No.12375086,  No.12335003, and National Key Research and Development Program of China under Contract No. 2020YFA0406400, and the Fundamental Research Funds  for the Central Universities under the Grant
	No. lzujbky-2024-oy02 and lzujbky-2023-it12.

\end{document}